\documentclass[prl,aps,groupedaddress,twocolumn]{revtex4}

\usepackage{graphicx}
\usepackage{amsmath}
\usepackage{amssymb}
\usepackage{bm}
\newcommand{\oper}[1]{\boldsymbol{\mathsf{#1}}}
\newcommand{\ket}[1]{\ensuremath{\left|{#1}\right\rangle}}
\newcommand{\bra}[1]{\ensuremath{\left\langle{#1}\right |}}
\newcommand{\bvec}[1]{{\bm{#1}}}
\newcommand{\brm}[1]{\bm{\mathrm{#1}}}
\begin{document}


\title{Quantum key distribution with higher-order alphabets using spatially-encoded qudits}


\author{S. P. Walborn}
\email[]{swalborn@if.ufrj.br}
\affiliation{Instituto de F\'{\i}sica, Universidade Federal do Rio de
Janeiro, Caixa Postal 68528, Rio de Janeiro, RJ 21941-972, Brazil}
\author{D. S. Lemelle}
\affiliation{Instituto de F\'{\i}sica, Universidade Federal do Rio de
Janeiro, Caixa Postal 68528, Rio de Janeiro, RJ 21941-972, Brazil}
\author{M. P. Almeida}
\affiliation{Instituto de F\'{\i}sica, Universidade Federal do Rio de
Janeiro, Caixa Postal 68528, Rio de Janeiro, RJ 21941-972, Brazil}
\author{P. H. Souto Ribeiro}
\affiliation{Instituto de F\'{\i}sica, Universidade Federal do Rio de
Janeiro, Caixa Postal 68528, Rio de Janeiro, RJ 21941-972, Brazil}


\date{\today}

\begin{abstract}
We present a proof of principle demonstration of a  quantum key distribution scheme in higher-order $d$-dimensional alphabets using spatial degrees of freedom of photons.  Our implementation allows for the transmission of $4.56$ bits per sifted photon, while providing improved security: an intercept-resend attack on all photons would induce an average error rate of $0.47$.  Using our system, it should be possible to send more than a byte of information per sifted photon.   
\end{abstract}

\maketitle
Though quantum key distribution (QKD) has become a commercial reality \cite{stix05}, there is still 
much interest in fundamental research.  One topic of fundamental importance is the design of protocols and implementations which increase the bit transmission rate and/or the security of the QKD scheme.  It has been pointed out recently that one can achieve both of these objectives by increasing the dimensionality of the system, that is, encoding a random key string  in $d$-dimensional qudits instead of the usual binary qubits \cite{bechmann00a,bourennane01}.    
\par
It is straightforward to generalize the well-known BB84 protocol \cite{bb84} to qudits \cite{bechmann00a,bourennane01,cerf02}, for  which it is possible to send on average $\log_{2}d$ bits per sifted qudit.    Higher-dimensional qudits are advantageous not only for an increased bit transmission rate, but also increased security. An eavesdropper employing an intercept-resend strategy would induce a qudit error rate of $E_{d}=\frac{1}{2}\frac{d-1}{d}$, since half the time she measures in the wrong basis, and consequently sends the wrong state with a probability of $(d-1)/d$ \cite{bechmann00a,bourennane01}.  
 \par
Experimentally, there are several methods of encoding $d$-dimensional qudits in photons, including time-bin \cite{bechmann00a}, orbital angular momentum \cite{leach02}, the polarization state of more than one photon \cite{bogdanov04}, and more recently position and linear momentum of entangled photons \cite{neves05,hale05}.  
\par
\begin{figure}
\includegraphics[width=7cm]{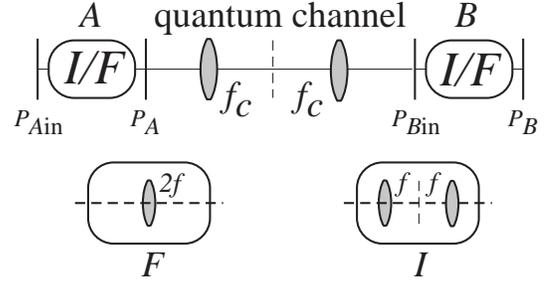}
\caption{\label{fig:basic} Illustration of QKD using imaging ($I$) and Fourier ($F$) optical systems.}
\end{figure}

Here we provide an experimental demonstration of quantum key distribution using higher-order $d$-dimensional alphabets encoded in the transverse spatial profile of single photons.  Our scheme is based on the standard BB84 protocol \cite{bb84}, in which Alice chooses which state to send based on the value of a random bit $a_1$, while her choice of basis is selected using random bit $a_2$.  A two-basis BB84 protocol using qudits works the same way \cite{bechmann00a,bourennane01}, however, Alice sends states according to the value of a random $d$-level ``dit".  A simple  illustration of our scheme is shown in FIG. \ref{fig:basic}.  Let us first discuss the choice of basis.  In our scheme, Alice ($A$) and Bob ($B$) encode (Alice) and decode (Bob) information in the transverse profile of single photons by choosing randomly between optical imaging systems and optical Fourier transform systems.       
   In order to avoid the quadratic phase factors that generally appear in an imaging system \cite{goodman96}, it is necessary to use a telescopic lens system, consisting of two confocal lenses.  This is equivalent to applying the Fourier transform operation twice, so that, as part of the protocol, Alice and Bob will each choose randomly between a single or double Fourier transform lens system.  For simplicity, let us assume that Alice and Bob use identical imaging systems, consisting of two lenses with focal length $f$, as well as identical Fourier systems consisting of a single lens with focal length $2f$.  The ``quantum channel" consists of a telescopic lens system consisting of two lenses with focal length $f_c$ which transmits Alice's output to Bob's input.                  
\par
In the following we will assume that the input field is a single photon state, which in the paraxial approximation can be described by 
 \begin{equation}
 \ket{\psi}=\int v(\brm{q}) \ket{\brm{q}}d\brm{q},
 \label{eq:1}
 \end{equation}
where $v(\brm{q})$ is the angular spectrum defined by
  \begin{equation}
  v(\brm{q}) = \int \mathcal{W}(\bvec{\rho},0) e^{-i\brm{q}\cdot\bvec{\rho}} d\bvec{\rho},
  \label{eq:2}
  \end{equation}   
  and  $\mathcal{W}(\bvec{\rho},0)$ is the input field at $z=0$ (plane $P_{A\mathrm{in}}$).  Here $\bvec{q}$ is the transverse component of the wave vector and $\bvec{\rho}$ is the transverse position coordinate.  The detection probability in plane $P_{B}$ for a given combination of lens configurations is given by $\mathcal{P}_{\alpha\beta}(\bvec{\rho})=|\mathcal{A}_{\alpha\beta}(\bvec{\rho})|^2$, where $\mathcal{A}(\bvec{\rho})=\bra{\mathrm{vac}}\oper{E}_{\alpha\beta}^{+}(\bvec{\rho})\ket{\psi}$ is the detection amplitude, $\oper{E}^{+}_{\alpha\beta}(\bvec{\rho})$ is the field operator for the entire lens system \cite{mandel95,walborn05d}, and $\alpha,\beta=I,F$ denotes either imaging or Fourier configurations.  For a series of $n$ confocal lenses, $\oper{E}^{+}_{\alpha\beta}(\bvec{\rho})$ simplifies to 
   \begin{align}
   \oper{E}^{+}_{\alpha\beta}(\bvec{\rho}) = & \mathcal{E} \int d\bvec{q}\int d\bvec{q}_{1}\cdots\int d\bvec{q}_{n} \oper{a}(\bvec{q}_{n}) e^{i\bvec{q}\cdot\bvec{\rho}} \times \nonumber \\
   & e^{-i\frac{f_{1}}{k}\bvec{q}_{1}\cdot\bvec{q}}\cdots
   e^{-i\frac{f_{n}}{k}\bvec{q}_{n}\cdot\bvec{q}_{n-1}},
   \label{eq:E}
   \end{align}
   where $\mathcal{E}$ is a constant, $k$ is the magnitude of the wavevector, $f_{j}$ is the focal length of the $j^{\mathrm{th}}$ lens and $\oper{a}(\brm{q})$ is the usual destruction operator.  
For the four possible lens systems illustrated in FIG. \ref{fig:basic}, the detection amplitudes are
  \begin{equation}
  \mathcal{A}_{FF}(\bvec{\rho}) = \frac{\mathcal{E}k^2}{2 f_c f}\mathcal{W}(\bvec{\rho},0),
  \label{eq:AFF}
  \end{equation} 
    \begin{equation}
  \mathcal{A}_{II}(\bvec{\rho}) = \frac{\mathcal{E}k^3}{f_c f^2}\mathcal{W}(-\bvec{\rho},0),
  \label{eq:AII}
  \end{equation}
  \begin{equation}
    \mathcal{A}_{IF}(\bvec{\rho}) = \frac{\mathcal{E}k^3}{2 f_c f^2}v\left(\frac{k}{2f}\bvec{\rho}\right),
  \label{eq:AIF}
  \end{equation}        
  and
   \begin{equation}
    \mathcal{A}_{FI}(\bvec{\rho}) = \frac{\mathcal{E}k^3}{2 f_c  f^2}v\left(\frac{k}{2f}\bvec{\rho}\right).
  \label{eq:AFI}
  \end{equation}          
 In our scheme, Alice encodes information into the input field by positioning an aperture $A(\bvec{\rho}-\bvec{\rho}_d)$  in plane $P_{A\mathrm{in}}$,  such that each aperture position $\bvec{\rho}_d$ corresponds to a character in the $d$-dimensional alphabet.  Assuming that the incident field is a plane wave, the input field is equivalent to the aperture function: $\mathcal{W}(\bvec{\rho},0)=A(\bvec{\rho}-\bvec{\rho}_d)$.    Eqs. (\ref{eq:AFF}) and (\ref{eq:AII}) show that when Alice and Bob choose the same lens configuration, Bob's detection amplitudes will reproduce the aperture function, and Bob should decode the correct character.  For complementary lens configurations the detection amplitudes are given by Eqs. (\ref{eq:AIF}) and (\ref{eq:AFI}), and are proportional to the Fourier transform of the aperture.  A well known property of the Fourier transform is that a shift in position space manifests as a phase in the Fourier transform ($\mathcal{F}$) space: $\mathcal{F}[A(\bvec{\rho}-\bvec{\rho}_d)]=\exp(ik\bvec{\rho}\bvec{\rho}_d/2f)\times \mathcal{F}[A(\bvec{\rho})]$.   Thus the detection probabilities $\mathcal{P}_{IF}$ and $\mathcal{P}_{FI}$ contain no information concerning the aperture position $\bvec{\rho}_d$. Even though Alice and Bob discard these results as part of the BB84 protocol, it is important that no information is available, as this guarantees that an eavesdropper cannot obtain information without causing an increase in the error rate.  
\par
   \begin{figure}
\includegraphics[width=8.3cm]{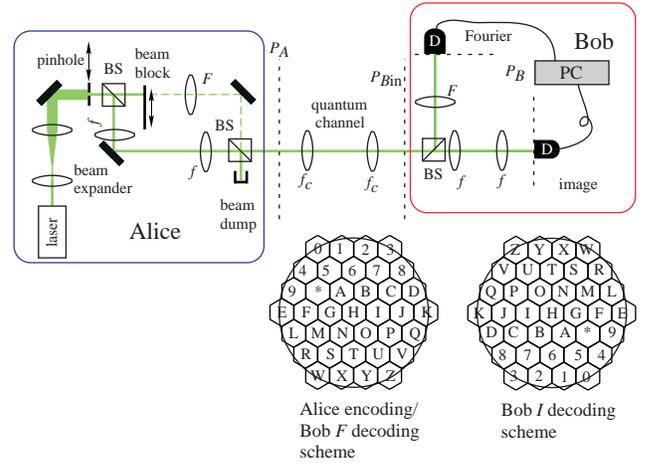}
\caption{\label{fig:setup} Experimental setup.}
\end{figure}
\par
FIG. \ref{fig:setup} shows the setup for an experimental demonstration of QKD using spatially encoded qudits.  As is common in most QKD implementations, our experiment was performed with an attenuated laser beam, which, though there are zero- and multi-photon terms present, can be used to  approximate a single photon state \cite{gisin02}.  The attenuated beam from a Coherent Verdi V5 laser (514 nm) was expanded by a factor of 4 using a beam expander consisting of $25$\,mm and $100$\,mm focal length lenses.  Information was encoded into the spatial profile by positioning a $200\,\mu$m  pinhole in Alice's transverse plane $P_{A\mathrm{in}}$.  The pinhole was mounted on a manual $x-y$ translation stage, though in principle a randomly-driven mechanical device could be used.  In order to implement both imaging and Fourier configurations, we constructed a Mach-Zehnder interferometer using 50-50 beam splitters (BS), in which one arm contained a telescopic imaging system ($f=100$\,mm), while the other contained a $200$\,mm focal length lens in a Fourier configuration.  To switch between imaging and Fourier configurations, we toggled manually between the two arms of the interferometer.  As interference is not actually used in the QKD scheme, the interferometer functions merely as a router.  However, the interference is useful for  initial alignment.  Pinholes were placed in the focal planes of the imaging and Fourier lenses in order to filter higher spatial frequencies.  As a result, the aperture function $A(\bvec{\rho}-\bvec{\rho}_d)$ can be approximated by a Gaussian.  The quantum channel consisted of a telescopic lens system ($f_{c}=150$\,mm). 
\par
 Using a BS, Bob chose randomly between imaging and Fourier systems.  His optical systems were identical to Alice's.  One single photon detector (equipped with $200 \mu$m diameter circular detection aperture and $\sim250$ nm bandwidth filter) was scanned throughout the Fourier detection plane, and one throughout the image detection plane.  Ideally, the detection system would consist of either two-dimensional multi-detector arrays, or CCD cameras with single-photon sensitivity \cite{abouraddy01b}.       
 \par
   The dimension $d$ of Alice and Bob' s alphabet is determined by the size of the aperture $A(\bvec{\rho})$ and its Fourier transform.  Alice and Bob must decide on the best way to define positions in transverse planes $P_{A\mathrm{in}}$ (Alice's aperture) and $P_B$ (Bob's detector) that will correspond to the characters in their alphabet.  To use the area available in the most efficient manner, we chose to approximate Alice's circular aperture and Bob's circular detection aperture with a hexagon (center to vertex distance $200 \mu$m).  Using this method, we were able to work with  a 37-dimensional (``septrigesimal") alphabet.  Alice  and Bob's encoding/decoding scheme is shown at the bottom of FIG. \ref{fig:setup}.   The circle corresponds to the area containing 99\% of the large Gaussian profile obtained using complementary $IF$ or $FI$ configurations.      
   \begin{figure}
\includegraphics[width=6cm]{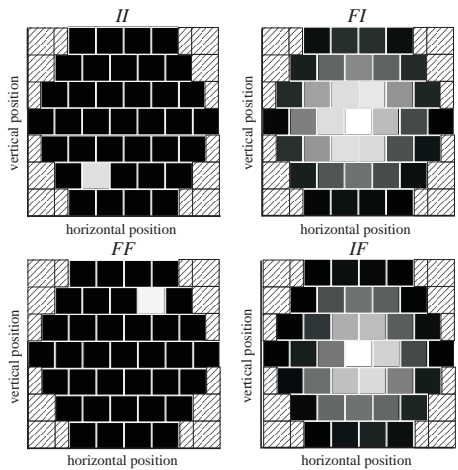}
\caption{\label{fig:letter7} Intensity distributions at Bob's detection plane for the four lens configurations $II$, $IF$, $FI$ and $FF$ for the case when Alice sends the character ``7".  Here lighter squares correspond to a larger number of photo-counts.}
\end{figure}
\par
FIG. \ref{fig:letter7} shows the intensity pattern at Bob's detection plane for the four possible lens configurations when Alice sends the character ``7".  The distributions were obtained by placing the detector at each of the pre-defined detection positions, so that each of the 37 squares in the figures correspond to a character in the alphabet.  For $II$ and $FF$ configurations, Bob detects the character ``7" with high probability, while for $IF$ and $FI$ configurations, he obtains a widened (Gaussian) distribution, which provides little information about the character Alice sent.       
   \begin{figure}
\includegraphics[width=8.3cm]{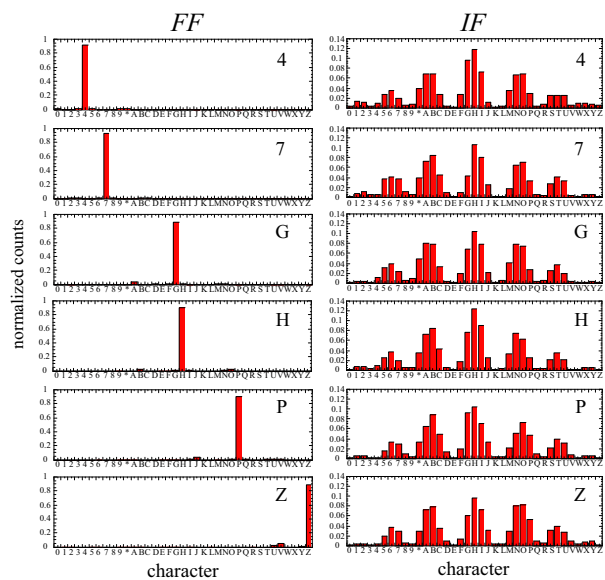}
\caption{\label{fig:Bp} Normalized counts for Bob's Fourier ($F$) detection system when Alice uses Fourier (left) and Image (right) encoding.}
\end{figure}
 \par
\begin{figure}
\includegraphics[width=8.3cm]{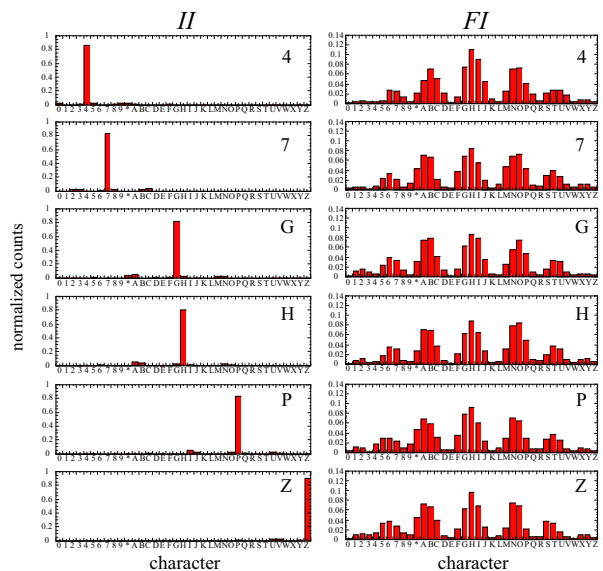}
\caption{\label{fig:Bx} Normalized counts for Bob's Image ($I$) detection system when Alice uses Image (left) and Fourier (right) encoding.}
\end{figure}

 As a better visualization of our results, 
 FIG.'s \ref{fig:Bp} and \ref{fig:Bx} show probability distributions as a function of each character for Bob's Fourier and image detection systems, respectively.   
  In both FIG.'s, Alice has sent the characters ``4", ``7", ``G", ``H", ``P" and ``Z".    When Bob uses the same lens configuration as Alice (left side in both figures), he detects the correct character with a high probability.  We obtained error rates $\mathcal{D}^{FF}_{k} \sim 0.06-0.11$ for the $FF$ configuration and $\mathcal{D}^{II} \sim 0.10-0.19$ for $II$ configuration.  Roughly 25\% of the error was due to photo-counts caused by unwanted ambient light and dark counts ($\sim$ 200 counts/sec), while the rest is due to misalignment and erroneous counts due to the hexagon pattern.  Using narrow band interference filters and detectors with a reduced dark count rate ($\sim$ 25-50 counts/sec), we estimate that the error rates could easily be reduced to about 5 - 15\%.  Further methods to reduce the $II$ and $FF$ error rate involve ``decoy" alphabet states and   will be discussed elsewhere \cite{walborn05d}.   
  \par
 FIG.'s \ref{fig:Bp} and \ref{fig:Bx} also show the results when Alice and Bob use conjugate $IF$ or $FI$ configurations, from which it can be seen that the detection probabilities $\mathcal{P}_{IF}$ and $\mathcal{P}_{FI}$ are the approximately the same for all character's sent by Alice.  We note that Bob's detection positions were defined according to the two-dimensional detection scheme shown in FIG. \ref{fig:setup}, so the several peaks shown in  the $IF$ and $FI$ patterns are actually slices of a 3D Gaussian distribution.   There is a difference between our QKD implementation and others:  the detection probabilities for complementary measurements are not constant for all states: $\mathcal{P}_{IF} = \mathcal{P}_{FI} \neq 1/{d}$ and thus the sifted key is not completely random.  However, after sifting, Alice and Bob can discard some of their results in order to obtain a completely random key string.  
  \par
  In order to minimize Eve's information, Alice should choose characters based on the distributions $\mathcal{P}_{IF}$ and $\mathcal{P}_{FI}$.  Suppose that Alice sends each character $k$ with probability $P_{k}$, obtained by averaging the $IF$ and $FI$ detection results.  The amount of information that can be sent from Alice to Bob is given by the Shannon information \cite{gisin02,bourennane01}, which in our case is
\begin{align}
I^{AB}=& I^A+\sum_{k=0}^{d-1}P_{k} (1-\mathcal{E}_k)\log_{2}(1-\mathcal{E}_{k}) \nonumber \\
& +\sum_{k=0}^{d-1} \sum_{j=0,j\neq k}^{d-1}\frac{P_k \mathcal{E}_kP_j}{1-P_k}\log_{2}\frac{\mathcal{E}_kP_j}{1-P_k}, 
\end{align}
where $\mathcal{E}_k$ is the error probability and $I^A=-\sum_{k=0}^{d-1}P_{k}\log_{2}P_{k}$ = 4.56 bits/photon is the information transmission in the absence of errors.  Our experimental error rates $\mathcal{D}^{II}$ and $D^{FF}$ varied between $0.06$ and $0.19$, giving $3.00 \leq I^{AB} \leq 3.96$  bits/photon.  For an intercept-resend attack on a fraction $\eta$ of the photons, the error rate is $\mathcal{E}_{k}=\frac{\eta}{2}(1-P_{k})$, which varies between $0.450 \eta$ and $0.499 \eta$.  In this case, Eve's information is given by $I^{E}=-\frac{\eta}{2} \sum_{k=0}^{d-1}P_{k}\log_{2}P_{k}=2.28 \eta$ bits/photon.  In order to employ classical error correction and privacy amplification, it is necessary that $I^{AB} > I^{E}$ \cite{gisin02}.   $I^{AB}=I^E = 1.858$ bits/photon occurs when the average error rate $\mathcal{E}=\sum_{k}P_k\mathcal{E}_k$ is about 0.38, much larger than our values of $0.06-0.19$.  
We note that the allowable error rate for cloning-based individual attacks on a two-basis $d=37$ protocol is 0.42\cite{cerf02} \footnote{We expect this limit to be slightly lower for our scheme, since the $IF$ and $FI$ results are not completely random.}.  
 \par
 Let us briefly discuss an important security issue particular to this implementation.  
 A more detailed security analysis will be provided elsewhere \cite{walborn05d}.
 In order for the transmission to be secure,  an eavesdropper Eve should not be able to determine when Alice is using the imaging or Fourier system to encode information.  If there exist detection positions at which Eve can detect photons that \textit{probably} correspond to an $I$-$F$ or $F$-$I$ (Alice-Eve) configuration, then she can deduce that she measured in the wrong basis, and choose not to resend the photon.  Eve's presence would then be marked only as the loss of a photon, and not a registered error.  In order to avoid this situation,  Alice and Bob must define their alphabet so that every detection position with a nonzero $IF$ or $FI$ detection probability also has a nonzero $II$ or $FF$ detection probability.   In this fashion, Eve cannot deduce whether she is measuring in the same basis as Alice or not.  On the other hand, if Eve can deduce that she probably measured in the correct basis, she gains nothing by not sending the photon.  Of course she has gained information and left no disturbance, but Alice and Bob can minimize these cases by  removing these characters from the final sifted key string, at the cost of a reduction in the size $d$ of  the alphabet. 
\par
   We have presented a proof of principle demonstration of QKD using spatially encoded qudits.  Generalization of our scheme to even larger dimensions is straightforward.  Using an even smaller aperture, it should be possible to encode an extremely large amount of information, increasing both the transmission rate as well as the security of the QKD protocol.  For example, using a $60 \mu$m pinhole, should give an alphabet of roughly $400$ characters in each photon, resulting in a transmission capacity of more than 1 byte per sifted photon.  
In terms of a real-world application, QKD based on spatial qudits seems best suited for free-space transmission as opposed to optical fibers.  In a free-space setup, disturbances in the wavefront due to propagation through the atmosphere might be monitored using a reference beam, and then corrected. 
  \begin{acknowledgments}
The authors acknowledge financial support from the Brazilian Millennium Institute for Quantum Information, CNPq, CAPES, FAPERJ, FUJB and PRONEX.  
\end{acknowledgments}

\end{document}